\documentclass{PoS}

\title{\[ \vspace{-4cm} \]
\noindent\hfill\hbox to 1.5in{\rm  } \vskip 1pt
\noindent\hfill\hbox to 1.5in{\rm SLAC-PUB-13541 \hfill  } \vskip
1pt \noindent\hfill\hbox to 1.5in{\rm June 2, 2004 \hfill}\vskip
40pt%
Quantum Calisthenics:  Gaussians, The Path Integral and Guided Numerical Approximations }

\ShortTitle{Quantum Calisthenics}

\author{\speaker{Marvin Weinstein}
\thanks{The animations referred to in the text can be viewed by going to $http:\backslash\backslash slac.stanford.edu
\backslash\sim niv $ and following the link to the animations for Quantum Calisthenics}\\
        SLAC National Accelerator Laboratory\\
        E-mail: \email{niv@slac.stanford.edu}}

\abstract{It is apparent to anyone who thinks about it that, to a large
degree, the basic concepts of Newtonian physics are quite intuitive,
but quantum mechanics is not.
My purpose in this talk is to introduce you to a new, much more intuitive
way to understand how quantum mechanics works.  I refer to this method
as a guided numerical approximation scheme and it is based upon a new
look at what the path integral tells us about states in Hilbert space.
I begin with simple exactly solvable models and show how to handle
problems which cannot be dealt with analytically, this includes the
treatment of the evolution of a Gaussian wave-packet in an anharmonic potential
as well tunneling problems (i.e., instanton effects)}

\FullConference{LIGHT CONE 2008 Relativistic Nuclear and Particle Physics\\
         July 7-11  2008\\
         Mulhouse, France}

\begin{document}

\def\ket#1{\vert #1 \rangle}
\def\bra#1{\langle #1 \vert}
\def\bracket#1#2{\langle #1 \vert #2 \rangle}
\def\vev#1{\left< #1 \right>}
\def\be{\begin{equation}}
\def\ee{\end{equation}}
\def\bea{\begin{eqnarray}}
\def\eea{\end{eqnarray}}
\def\x{{\bf x}}
\def\p{{\bf p}}
\def\H{{\bf H}}
\def\N{{\bf N}}
\def\V{{\bf V}}

\section{Quantum Mechanics Isn't Intuitive !}

It is apparent to anyone who thinks about it that, to a large
degree, the basic concepts of Newtonian physics are quite intuitive,
but quantum mechanics is not.
My purpose in this talk is to introduce you to a new, much more intuitive
way to understand how quantum mechanics works.

I begin with an incredibly easy way to
derive the time evolution of a Gaussian wave-packet for
the case free and harmonic motion without any need to know the
eigenstates of the Hamiltonian.  This discussion
is completely analytic and I will later use it to relate the solution
for the behavior of the Gaussian packet to the Feynman path-integral
and stationary phase approximation.  It will be clear that using
the information about the evolution of the Gaussian in this way
goes far beyond what the stationary phase approximation tells us.

Next, I introduce the concept of the {\it bucket brigade approach\/}
to dealing with problems that cannot be handled totally analytically.
This approach combines the intuition obtained in the
initial discussion, as well as the intuition obtained from the path-integral,
with simple numerical tools.  My goal is to show that, for any specific
process, there is a simple Hilbert space interpretation of the
stationary phase approximation. I will then argue that, from the
point of view of numerical approximations, the trajectory obtained from
my generalization of the stationary phase approximation specifies that
subspace of the full Hilbert space that is needed to compute the time
evolution of the particular state under the full Hamiltonian.  The
prescription I will give is totally non-perturbative and we will see,
by the grace of Maple animations computed for the case of the anharmonic
oscillator Hamiltonian, that this approach allows surprisingly accurate
computations to be performed with very little work.  To view these
animations go to $http:\backslash\backslash slac.stanford.edu\backslash \sim niv$.
I think of this approach to the path-integral as defining what I call a
{\it guided numerical approximation scheme\/}.

After the discussion of the anharmonic oscillator I will turn to tunneling
problems and show that the {\it instanton\/} can also be though of in
the same way.  I will do this for the classic problem of a double well
potential in the extreme limit when the splitting between the two lowest
levels is extremely small and the tunneling rate from one well to
another is also very small.

\section{Gaussian Wavefunctions and the Path Integral}

Discussions of one-dimensional quantum mechanics
usually begin by considering the space
of square integrable functions $\psi(x)$ on the interval $-\infty < x < \infty$.
This space of functions
is acted upon two operators, ${\bf p}$ and ${\bf x}$, where the action of these operators
on a state $\psi(x)$ is defined to be
\bea
         {\bf x}\, \psi(x) = x \psi(x) \\
         {\bf p}\, \psi(x) = {1 \over i} {d \psi \over dx}(x);
\eea
i.e., ${\bf x}$ is simply multiplication by the variable $x$ and ${\bf p}$ is
differentiation with respect to the variable $x$.
Given these definitions it is simple to show that
${\bf x}$ and ${\bf p}$ satisfy the commutation relation
\be
   [ \x, \p ] = i.
\ee

In what follows we will devote a great deal of attention to the properties
of Gaussian wave-packets.  For our purpose a Gaussian packet of width $1/\sqrt{\gamma}$
is defined as the solution to the equation
\be
      ( i\,\p + \gamma \x ) \psi(x) = 0.
\label{Gaussdef}
\ee
To show this is the same as the usual condition is simple, but worth doing explicitly since we
will use it over and over again.  The steps are
\bea
    ( {d \psi \over dx}(x) + \gamma \psi(x)) = 0 \\
    \psi(x) = C e^{- {1 \over 2} x^2}
\eea
where constant $C$ is determined by the normalization condition
\be
     \int_{-\infty}^{\infty} \,dx\, \psi^{\ast}(x) \,\psi(x) = 1,
\ee
that means that $C = \left({\gamma \over \pi} \right)^{1/4}$.

From this point on we will write a Gaussian satisfying this equation by
the symbol $\ket{0_\gamma}$.

\section{Shifted Gaussians}

In what follows it will be important for us to consider Gaussian wave-packets centered
about points $\tilde{x} \ne 0$; i.e., Gaussian packets shifted away from the origin.
As is customary, these packets will be constructed by applying the operator
\be
    U(\tilde{x}) = e^{-i {\bf p} \tilde{x}} = \sum_{i=0}^\infty { (-i)^n \over n!} \p^n \,\tilde{x}^n
\ee
to the state $\ket{0_\gamma}$ to obtain the state $\ket{\tilde{x}_\gamma}$; where $\tilde{x}$ is an arbitrary number and $\p$ is the
momentum operator.

To see that this operation does what we want observe that
\bea
    e^{-i \p \tilde{x}}\, \x e^{i \p \tilde{x}}  &=& \x - i\,\tilde{x} [\p ,\x ] + {(-i)^2\over 2!}\,
    \tilde{x}^2\, [ \p, [\p,\x]] + \ldots \\
    &=& \x - \tilde{x}
\eea
where all but the first two terms of the expansion vanish since the commutator of $\p$ with
$\x$ is proportional to the unit operator.
Thus,
\bea
    e^{-i \p\, \tilde{x}} \, (i \p + \gamma\, \x )\, \ket{0_\gamma} &=& 0 ,\\
    e^{-i \p\, \tilde{x}} \, (i \p + \gamma\, \x )\, e^{i \p\, \tilde{x}}\, (e^{-i \p\, \tilde{x}} \, \ket{0_\gamma}) &=& 0 , \\
    e^{-i \p\, \tilde{x}} \, (i \p + \gamma\, \x )\, e^{i \p\, \tilde{x}}\, \ket{\tilde{x}_\gamma} &=& 0 .
\label{shifted}
\eea
Now observe that $\p $ commutes with $U(\tilde{x})$ so that Eq.\ref{shifted} becomes
\be
    ( i \p + \gamma (x - \tilde{x}) ) \ket{\tilde{x}_\gamma} = 0 .
\ee
Using the previous argument, thinking of $\ket{\tilde{x}_\gamma}$ as a function
of $x$, we have
\be
   \ket{\tilde{x}_\gamma} = \left( {\gamma \over \pi} \right)^{1/4}\, e^{-{\gamma \over 2}\,(x-\tilde{x})^2}.
\ee
This, of course, is what we wished to show.

I have only included this elementary discussion to show how one can manipulate the equation
that defines a Gaussian packet in order to obtain a useful result.  In the next section I
will use the same sort of argument to derive the evolution in time of an arbitrary
Gaussian packet.

\section{Evolution of a Gaussian With Free Hamiltonian}

The operator form of the time-dependent Schrodinger equation, given a Hamiltonian ${\bf H}$,
says that the state $\ket{\psi(t)}$ satisfies the equation
\be
    {d \over dt} \ket{\psi(t)} = -i {\bf H}\ \ket{\psi(t)} .
\ee
As is well known, the solution to this equation is
\be
    \ket{\psi(t)} = e^{-i\,t\,{\bf H}}\  \ket{\psi(t)} .
\ee
To derive the time evolution of a Gaussian packet we
multiply Eq.~\ref{Gaussdef} by $e^{-i\,t\,H}$, where $H$ is\
defined to be
\be
    H = {\p^2 \over 2 m},
\ee
for some mass {\it m}; i.e., we consider
\bea
    e^{-i\,t\,{\p^2 \over 2 m }}\ ( i\, \p + \gamma\, \x )\ \ket{0_\gamma} &=& 0\\
    e^{-i\,t\,{\p^2 \over 2 m }}\ ( i\, \p + \gamma\, \x )\ e^{ i\,t{\p^2 \over 2 m}} \ \ket{0_\gamma(t)} &=& 0 \\
    ( i\,\p(t) + \gamma\, \x(t) )\ \ket{0_\gamma(t)} &=& 0 \label{ketoftone};
\eea
where the time dependent operators $\x(t)$ and  $\p(t)$ are
\bea
    \x(t) &=& e^{-i\,t\ {\p^2 \over 2 m} }\ \x \ e^{i\,t\, {\p^2 \over 2 m}} = \x - t\,{\p \over m}, \\
    \p(t) &=& e^{-i\,t\ {\p^2 \over 2 m} }\ \p\ e^{i\,t\, {\p^2 \over 2 m}} = \p.
\label{xoftpoft}
\eea
These results follow from the definition of the exponential and the commutation relations.
Substituting this into Eq.~\ref{ketoftone} we obtain
\bea
     ( i\,\p + \gamma\,( \x - {t\over m}\, \p) \ket{0_\gamma(t)} &=& 0 \\
    \left( \left(1 +  {i \gamma\, t \over m}\right)\,i\, \p
    + \gamma \, \x \right) \ \ket{0_\gamma(t)} &=& 0 \\
    ( i\,\p + \gamma\,(t) \x ) \ \ket{0_\gamma(t)} &=& 0 .
\eea
This, as we have already shown, means
\be
    \ket{0_\gamma(t)} = C(t)\ e^{-{1 \over 2}\,\gamma(t)\, x^2},
\ee
where
\be
    \gamma(t) = { \gamma \over 1 + i {\gamma\,t \over m}} .
\ee
The fact that
\be
    C(t) = \left(\gamma \over \pi \right)^{1/4} \ {1 \over \sqrt{ 1 + i {\gamma\,t \over m}}}
\ee
follows directly from the equation
\be
    {d \over dt} ( C(t) e^{-{1 \over 2}\,\gamma(t)\,x^2} ) = -i\, C(t)\,{ d^2 \over dx^2}
e^{-{1 \over 2}\,\gamma(t)\,x^2} ),
\ee
which is simply a differential equation for $\ln(C(t))$.
I have bothered to include the entire derivation of the time-dependent wave-function
to show how powerful manipulating the defining equation for the Gaussian packet can
be.  Because the time dependent packet drops off as $e^{-{1 \over 2}\, \gamma(t) x^2}$
where $\gamma(t)$ is complex, we will refer to this as a {generalized Gaussian\/} packet.
An animation showing how this wave-function looks as it evolves in time is shown at
the beginning of the html file that I referred to previously this paper.  It is followed by
animations showing how different coherent states evolve using the free Hamiltonian.

\section{The Harmonic Oscillator}

Now that we have the exact solution of the time evolution of an arbitrary
Gaussian packet under the free Hamiltonian, let us consider the next simplest case,
the harmonic oscillator Hamiltonian; i.e.
\be
    H = {\p^2 \over 2\,m } + { m\,\omega^2 \over 2} \x^2 .
\ee
Once again if we start by multiplying the defining equation of a Gaussian packet
by the exponential of $H$ we find
\be
    e^{- i\,t\,H} \ ( i\,\p + \gamma\,\x)\ \ket{0_\gamma} = 0 ,
\ee
or
\be
    ( i\,\p(t) + \gamma\,\x(t)) \ket{0_\gamma(t)} = 0,
\label{harmonicwvfn}
\ee
where the time-dependent operators $\p(t)$ and $\x(t)$ are defined to be
\be
    \p(t) = e^{-i\,t\,H}\ \p\ e^{i\,t\,H} \qquad {\rm and\ }  \qquad
    \x(t) = e^{-i\,t\,H}\ \x\ e^{i\,t\,H} .
\ee
It follows immediately from this equation and the commutation relations of $\p$ and $\x$
that
\be
    {d \x \over dt} = {1 \over m} \p(t) \qquad {\rm and } \qquad
    {d \p \over dt} = - \omega^2\, \x(t) .
\ee
Since this is simply a first order differential equation with the boundary conditions
$\p(t=0) = \p$ and $\x(t=0) = \x$, it has the unique solution
\bea
    \x(t) &=& \cos(\omega\,t)\ \x + {1 \over m\,\omega} \sin(\omega\,t) \p \\
    \p(t) &=& \cos(\omega\,t)\ \p - m\,\omega \sin(\omega\,t) \x
\eea
Substituting these into Eq.~\ref{harmonicwvfn} we obtain, as before,
\be
    \left( i\,\p + \gamma\,\left( {\cos(\omega\,t) + { i\,m\,\omega \over \gamma}
    \ \sin(\omega\,t) \over
    \cos(\omega\,t) + { i\,\gamma \over m\,\omega}
    \ \sin(\omega\,t) } \right)\ \x \right)\,\ket{0_\gamma(t)} = 0
\ee
Thus we see that the evolution of a Gaussian packet in a harmonic oscillator
potential is once again a generalized Gaussian with a  $\gamma(t)$ that is periodic
in time.  Note, if and only if $\gamma = m\,\omega$, the packet doesn't change in time and
it is simply multiplied by a phase $e^{-i\,t\,\omega / 2}$; i.e. when $\gamma = m\,\omega$
the packet is an eigenstate of the harmonic oscillator Hamiltonian.

\section{Coherent States}

Previously we discussed Gaussians shifted to a mean position $\tilde{x}$.  Now we will
generalize the shifted state to one that has a non-vanishing expectation value
for both $\x$ and $\p$; i.e., consider the time evolution of the state
\be
    e^{-i\,\p \,\tilde{x} + i\,\tilde{p}\,\x} \ \ket{0_\gamma}
\ee
This is called a coherent state.

Clearly, using the previous arguments, multiplying this state
to the left by $e^{-i\,t\,H}$ we obtain
\be
    e^{-i\,\p(t) \,\tilde{x} + i\,\tilde{p}\,\x(t)} \ \ket{0_\gamma(t)} .
\ee
Collecting terms this can be rewritten as
\be
    e^{-i\,\p \,\tilde{x}_{\rm class}(t) + i\,\tilde{p}_{\rm class}(t)\,\x} \ \ket{0_\gamma(t)} .
\label{topathinteg}
\ee
where $\ket{0_\gamma(t)}$ was calculated in the preceding section and
$\tilde{x}_{\rm class}(t)$ and $\tilde{p}_{\rm class}(t)$ are the solutions to the
classical equations of motion for a particle moving in a harmonic potential
that initially is located at the position $\tilde{x}$ with momentum $\tilde{p}$.
It is not an accident that this is the same trajectory one would obtain by
doing the stationary phase approximation to the path integral.  To make the analogy between
the formula for the propagation of an arbitrary shifted Gaussian
and the stationary phase approximation more striking let us rewrite Eq.~\ref{topathinteg}
as
\be
e^{i\,\tilde{p}_{\rm class}(t)\,\tilde{x}_{\rm class}(t)}\
e^{-i\,\p \,\tilde{x}_{\rm class}(t)
+ i\,\tilde{p}_{\rm class}(t)\,(\x - \tilde{x}_{\rm class}(t))} \ \ket{0_\gamma(t)} .
\ee
This form of the generalized Gaussian packet shows that the packet
center moves along the classical trajectory for a particle starting with the given
initial mean position and mean momentum.  Furthermore, it shows that, if we write the
position dependent phase factor that gives the shifted packet the correct mean classical
momentum, so that it has the value one at the packet center, then the entire
packet is multiplied by a time dependent phase factor that is the exponential of the
classical action.  These results are also what is seen in derivations of the path-integral
using coherent states with a single fixed value for $\gamma$.  What is not captured
in the coherent state derivation of the path-integral is the fact that $\gamma$ changes
in time and, in fact, becomes complex.  This is why that approach is
less powerful than what I will do next.

\section{The Bucket-Brigade Approach to the Path-Integral}

We now understand how a generalized Gaussian packet propagates with the free or harmonic
oscillator Hamiltonian. Now, let us spend a few moments connecting this knowledge to the
usual derivation of the path integral using Gaussian coherent states.
Most derivations begin by rewriting the time evolution
operator as a product of the evolution operator for many small time steps and inserting
a complete set of states between each term in the product;i.e.,
\be
   \bra{\gamma_{\rm final}}\, e^{-i\,t\,\H} \ket{\gamma_{\rm init}} =
    \bra{\gamma_{\rm final}}\,\ldots\,\ket{\gamma_{j+1}}\bra{\gamma_{j+1}} e^{-i\,t\,\H \, /n} \,
    \ket{\gamma_{j}}\bra{\gamma_{j}} e^{-i\,t\,\H \,/n} \,
\ket{\gamma_{j-1}}\bra{\gamma_{j-1}} \ldots \ket{\gamma_{\rm init}}.
\ee
In Feynman's derivation of the path integral this complete set
of states are $\delta$-functions of $x$ or $p$, as appropriate.
Later derivations used coherent states, since the shifted coherent
states form an over-complete basis in terms of which one can
construct a resolution of the identity operator. In either case,
after deriving this identity one customarily makes the {\it
stationary phase approximation\/}, which in effect selects a single
intermediate state at each step. In all of these cases, however,
even for the exactly solvable cases of the free particle, or the
harmonic oscillator, the states selected by this step are not good
approximations to the true evolution. We have already seen that the
correct evolution of a free particle or a particle in a harmonic
oscillator potential is a generalized shifted Gaussian with a
complex $\gamma(t)$. Obviously, if we insert these states as
intermediate states then the stationary phase approximation would
produce the exact answer. This observation almost brings us to the
formulation of the {\it bucket-brigade\/} approach to dealing with
the Schrodinger equation.  The missing step is the observation that
as the number of steps in the decomposition of the evolution
operator increases the number of states selected by the stationary
phase approximation increases too. However, since these states are
not orthogonal to one another, the number of significantly linearly
independent states doesn't grow in the same way. To be more precise,
I define the notion of significantly linearly independent states as
follows:  let the integers $M$ and $N$, with $ M < N$ define two
decompositions of the time interval in the decomposition. Let
$\ket{\gamma_i}$ and $\ket{\psi_j}$ be the two sets of states
defined by the corresponding stationary phase condition, i.e. let
them be the generalized Gaussian packets obtained by exactly
propagating the initial state a time $t/M$ or $t/N$. The larger set
of states will not be significantly linearly independent of the
smaller set if all of the larger states can be represented to some
pre-defined accuracy as a linear combination of the smaller set of
states.

Since dividing the time interval over which the evolution
is occurring into ever smaller slices does not lead to increasing
numbers of significantly linearly independent states, it
follows that one can describe the continuous time evolution of the
initial state to arbitrary accuracy be restricting attention to a
finite dimensional sub-space of the full Hilbert space.  In this
section I will show that this is the case for free and harmonic
evolution. To be precise, I will show that in order to compute the
states $e^{-i\,t\,\H} \ket{\gamma_0}$ for all values of $t$ between
some $t_{\rm initial}$ and $t_{\rm final}$ to high accuracy it
suffices to, given some discrete set of states $\ket{\psi_n}$,
compute the truncated operators
\be
    H_{n\, m} = \bra{\gamma_n} \H \ket{\gamma_m} \qquad {\rm and}  \qquad N_{n\,m} =
    \bra{\gamma_n} \N \ket{\gamma_m},
\ee
and then exponentiate the finite matrix $H_{n m}$ after transforming it
to the orthonormal basis defined by the $\ket{\psi_n}$'s.  In the
accompanying .html I show how well this works for various initial states
evolving either under the free Hamiltonian or the harmonic oscillator
Hamiltonian.

\section{A Non-Trivial Example: The Anharmonic Oscillator}

I have argued that the bucket-brigade idea says that, with
no significant loss of accuracy, we can restrict attention to a relatively
small subspace of Hilbert space to compute the continuous time
evolution of a given packet, I will now show that the same is true
for Hamiltonians for which the time evolution cannot be exactly
computed.  To show how this works let us begin by considering the
case of the anharmonic oscillator; i.e., the system defined by the
Hamiltonian
\be
    \H = {\p^2 \over 2\,m } + \lambda\,\x^4 .
\ee
We will start, as before, with a Gaussian packet defined by the equation
\be
    \left( i\,\p + \gamma\,\x \right) \,\ket{0_\gamma} = 0 .
\ee
Then, it follows that the time-evolved packet satisfies the equation
\be
    e^{-i\,t\,\H} \left( i \p + \gamma \x \right)\ket{\gamma_0} = 0 ,
\ee
or
\be
    \left( i \p(t) + \gamma \x(t) \right)\ket{\gamma(t)} = 0 ,
\ee
where $\x(t)$ and $\p(t)$ are defined as in Eq.~\ref{xoftpoft}.
It not possible, however, to compute $\x(t)$ or $\p(t)$ exactly.  We
can however non-perturbatively approximate the evolution of a state
\be
    \ket{x_i,p_i,\gamma(t)} = e^{-i\,x_i\,\p + i\,p_i\,\x}\,\ket{0_{\gamma(t)}}
\ee
by evolving it with the {\it effective quadratic Hamiltonian\/}
\bea
    \H_{\rm eff}( x_i, p_i, \gamma(t) ) &=& {(\p+p_i)^2 \over 2\,m}  +
    \lambda\, \bra{{\it Re}(\gamma(t))}\, \V(\x + x_i)\,\ket{{\it Re}(\gamma(t))}\\
    &+&\lambda\, \bra{{\it Re}(\gamma(t))}\,{d \over d\x} \V(\x + x_i)\,\ket{{\it Re}(\gamma(t))}\,(\x - x_i)\\
    &+&\lambda\, \bra{{\it Re}(\gamma(t))}\,{d^2 \over d \x^2} \V(\x + x_i)\,\ket{{\it Re}(\gamma(t))}\,
    (\x - x_i)^2 .
\eea
Note that in order to guarantee that the Hamiltonian is hermitian the expectation values are
computed for Gaussian packets where $\gamma(t)$ is replaced by a Gaussian with the same
where $\gamma(t)$ is replaced by the real part of $\gamma(t)$.  Thus, with this in
mind we define the iterative procedure where we begin with a generalized Gaussian
\be
    \ket{\psi_n} =   C_n\,e^{i\,p_n\,(x - x_n)}\,e^{-{1 \over 2}\,\gamma_n\,( x- x_n)^2}
\ee
and evolve it with a quadratic Hamiltonian with the generic form
\be
    \H_{\rm eff} = {(\p+p_n)^2 \over 2\,m} + V_n - F_n\,(x-x_n) + {m\,\omega^2_n \over 2}.
\ee
Applying the formulas we already derived for the evolution of a generalized
Gaussian in a harmonic potential we see that we get a new generalized Gaussian
of the form
\be
    \ket{\psi_{n+1}} = C_{n+1}\,e^{i\,p(\delta t)_{n+1}\,(x - (x_n + \tilde{x}(\delta t)_{n+1}))}\,
    e^{-{\gamma_{n+1} \over 2}\,(x - (x_n + \tilde{x}(\delta t)_{n+1}))^2}
\ee
where the quantities appearing in this equation are given by the formulas
\be
    C_{n+1} = C_n\,{1 \over \sqrt{\cos(\omega_n\,\delta t) + {i\,\gamma_n \over m\,\omega_n}\,
    \sin(\omega_n\, t)}}\,e^{i\,F_n^2 \over 2\,m\,\omega_n^2}\,e^{-i {F_n \over m\,\omega_n^2}\,
    (p(\delta t)_{n+1} - p_n)}\,e^{i\p(\delta t)_{n+1}\,x(\delta t)_{n+1}},
\ee
and
\bea
    x(\delta t)_{n+1} &=& {F_n \over m\,\omega_n^2} \left( 1 - \cos(\omega_n\,\delta t)\right)
    + {\p_n \over m\,\omega_n}\,\sin(\omega_n \delta t)  \\
    p(\delta t)_{n+1} &=& \cos(\omega_n\,\delta t)\,p_n + {F_n \over \omega_n}\,\sin(
    \omega_n\,\delta t) \\
    \gamma_{n+1} &=& { \cos(\omega_n\, \delta t) + {i\,m\,\omega_n \over \gamma_n}
    \,\sin(\omega_n\,\delta t) \over
     \cos(\omega_n\, \delta t) + {i\,\gamma_n \over m\,\omega_n}
    \,\sin(\omega_n\,\delta t)} .
\eea
Note, that at the $n^{th}$ step the Hamiltonian parameters, for the case
of the anharmonic oscillator are defined by the
equations
\bea
  V_n &=& \lambda \, x_n^4 + {3\,\lambda \over \gamma_n}\,x_n^4
    + {3\,\lambda \over 4\,\gamma_n^2} \\
    F_n &=& - 4\,\lambda\,x_n^3 - {6\,\lambda \over \gamma_n}\,x_n \\
    \omega_n &=& \sqrt{ {12\,\lambda \over m}\,x_n^2 + {6\,\lambda \over m\,\gamma_n}} .
\eea
One repeats this process again and again, recomputing the effective
Hamiltonian at each stage, and obtains a basis for the truncated Hilbert space.
To compute the continuous time evolution of the initial state we
exponentiate the truncated Hamiltonian, obtained by computing the
matrix elements of the exact Hamiltonian between these Gaussian packets.
Since these are Gaussians, this is easy to do.
The comparison of this approximate computation and
the exact result obtained from a brute force numerical approximation
is shown in the previously referred to html file.
That file contains many animations
showing how a free Gaussian packet evolves in time under various circumstances,
as well as how a Gaussian in a harmonic potential evolves in time.
It also contains animations that compare the evolution using discretized,
{\it bucket-brigade\/} states to exact solutions, both for the exactly
solvable systems, as well as for the anharmonic oscillator.
You will see that, in all cases, the agreement
between the real and imaginary part of the wave-fucntion
for approximate and exact calculation is quite remarkable.

\section{Tunneling and Instantons}

The final issue I want to touch upon is tunneling, which is important both
for problems related to tunneling between different minima of a potential,
and to general problems of scattering from a non-square barrier.

Consider the Hamiltonian for a particle in a double-well potential
\be
    \H = {1 \over 2\,m}\,\p^2 + \lambda\,(\x^2 - f^2)^2 .
\ee
If we now attempt to find stationary Gaussians there will be two
solutions, one in each well.  As before the classical moments, $\tilde{p}$ of each solution
must be set to zero.  Next, as is shown in the .html file, the condition
that the force, the expectation value of the derivative of the potential,
vanish means that in the left-hand well the Gaussian is shifted slightly to
the right of the minimum and in the right-hand well the Gaussian is shifted
slightly to the left.  The parameter $\gamma$ for each of the Gaussian's
doesn't evolve in time determines $\gamma$ in terms of the appropriate expectation value
of the second derivative of the potential.  If we stop at this point
the bucket-brigade approach would now determine the future behavior of the
system by computing the matrix elements of the Hamiltonian between these two
states and the metric formed by taking the overlaps of the two states.
This result alone tells us that tunneling takes place.  However, and that
is what we now wish to study, it severely underestimates the tunneling
rate when the mass is large and the wells are well separated.
The question is "what states do we have to add in order to compute the
tunneling rate accurately ?".
The answer is, of course, determined by the instanton calculation.

The key point is that in order to improve the energies of the two
lowest states it suffices to apply the operator $e^{- t\,\H}$ to
compute the effective Hamiltonian \cite{COREpaper}-\cite{COREpaper2}-\cite{somepapers}
\be
    \H_{i j}(t) = \left[\bra{\psi_i}\, e^{-t\,\H}\,\ket{\psi_l}\right]^{-1/2}\,
    \bra{\psi_l}\, e^{-{t\over 2}\,\H}\, \H \, e^{- {t\over 2}\,\H} \ket{\psi_m}\,
    \left[\bra{\psi_m}\, e^{-t\,\H}\,\ket{\psi_j}\right]^{-1/2}.
\ee
If we now follow the path integral procedure and divide the product up
into a number of steps and insert a set of generalized Gaussian packets
\be
    e^{-i\,\p\,\tilde{x}(t) + i\,\tilde{p}(t) \,\x}\ket{\gamma(t)}
\ee
where $\tilde{p}(t)$ is chosen so that, as is always the case,
\be
    \tilde{p}(t) = {1 \over m} \,{d \tilde{x}\over dt}(t),
\ee
and for simplicity $\gamma(t)$ is chosen to be a constant, then it follows that this
contribution to the transition element is maximized if the function $\tilde{x}(t)$
satisfies the equation
\be
    {d^2 \over dt^2}\tilde{x}(t) = 4\,\lambda\tilde{x}(t)\,\left(\tilde{x}(t)^2 - f^2 \right)
\ee
where $\tilde{x}(t) = -c_{\rm min}$ and $d \tilde{x}(t=0) / dt = 0$, and $\gamma$ is chosen
to be the same as that for the initial and final Gaussian packet. A picture of the solution
and the discretized choice of a finite number of these states is shown in the appropriate
section of the .html file.

If we assume, as before, that these are the correct states to use to compute the
time evolution of the initial Gaussian we simply compute the truncated Hamiltonian
and exponentiate it. The corresponding animations in the accompanying .html file show that this
computation is remarkably accurate for the real and
imaginary part of the wavefunction, as a function of time, as well as for the tunneling rate.

It is a straightforward matter to extend these ideas to field theory, however time and space
preclude discussing this question at this time.

\end{document}